\documentclass[showpacs,aps,preprint,prd,nofootinbib,floatfix,amsmath,showkeys,11pt]{revtex4-1}
\usepackage{graphicx} 
\usepackage{xcolor}
\usepackage{mathtools}
\usepackage{amsfonts,amssymb}
\usepackage{soul}
\usepackage{ulem}
\usepackage{cancel}
\usepackage{float}

\usepackage{subfigure}
\usepackage{bm}
\usepackage{slashed}

\newcommand{\anaes}[1]{}

\newcommand{\soul}[1]{}
\DeclareGraphicsExtensions{.pdf,.png,.jpg}
\newcommand{\sigmam}{$\langle\sigma_{ann} v_{rel}\rangle-m_\chi\,$}
\newcommand{\sigmav}{$\langle\sigma_{ann} v_{rel}\rangle\,$}

\usepackage{natbib}
\usepackage[pdftex]{hyperref}
\hypersetup{colorlinks=true,linkcolor=blue} 

\begin{document}
	
	\title{Signatures of Dipolar Dark Matter on Indirect Detection}

	\vspace*{-15mm}
	\begin{flushright}
		CIFFU 19-02
	\end{flushright}
	\vspace*{0.7cm}

\author{C. Arellano-Celiz$^{1,3}$, A. Avilez-L\'opez$^{1,3}$, J. E. Barradas-Guevara$^{1,3}$, A. Carrillo-Monteverde$^{2,3}$, J. L. D\'iaz-Cruz$^{1,3}$, O. F\'elix-Beltr\'an$^{2,3}$\footnote{
olga.felix@correo.buap.mx}} 
\affiliation{
	$^{1}$Facultad de Ciencias F\'{\i}sico Matem\'aticas, Benem\'erita Universidad Aut\'onoma de Puebla, Apdo. Postal 1152, Puebla, Pue., M\'exico.\\
	$^{2}$Facultad de Ciencias de la Electr\'onica, Benem\'erita Universidad Aut\'onoma de Puebla, Apdo. Postal 542, C.P. 72570, Puebla, Pue., M\'exico. \\
	$^{3}$Centro Internacional de F\'isica Fundamental (CIFFU), Puebla, Pue., C.P. 72570, M\'exico.
}

\date{\today}

\begin{abstract} 
 In this work we study the annihilation of fermionic dark matter, considering it as a neutral particle with non vanishing magnetic ($M$) and electric ($D$) dipole moments. Effective cross section of the process  $\chi \overline{\chi} \rightarrow \gamma \gamma$ is computed  starting from a general form of the coupling $\chi \overline{\chi} \gamma$ in the framework of an extension of the Standard Model. By taking into account annihilation of dark matter pairs into mono-energetic photons, we found that for masses of $O(10^2)$, an electric dipole moment $\sim 10^{-16}\, \textrm{e cm}$ is required to satisfy the current relic density inferences. Additionally, in order to pin down models viable to describe the physics of dark matter in the early Universe, we also constrain our model according to recent measurements of the temperature anisotropies of the  cosmic microwave background radiation, and report constraints to the electric and magnetic dipole moments for a range of masses within our model. 
\end{abstract}



\maketitle

\section{Introduction}\label{sec:introduction}
The enigma of dark matter (DM) is perhaps the  most interesting problem in modern astrophysics, moreover that it has led to the incursion of elementary particle physics~\cite{Zwicky:1933gu,Rubin:1962}. 
The joint work of these two disciplines has as one of its main objectives to determine the nature and properties of DM, either through direct or indirect detection. Nowadays, the evidences from galactic dynamics (rotation curves), galaxy clusters, structure formation, as well as Big Bang's nucleosynthesis and the Cosmic Microwave Background (CMB), suggest that baryons do not suffice to explain these observations; therefore, most of the non-relativistic missing matter prevailing in the Universe must be non-baryonic~\cite{Peebles:2017bzw,Roos:2012,Bergstrom:2000,Bertone:2005}. 

Thus, physics beyond the Standard Model (BSM) has been considered in order to accommodate a non-baryonic DM candidate~\cite{Jungman:1996,Bertone:2005,DiazCruz:2007be,DiazCruz:2007fc}.
 
Weakly interactive massive particles (WIMPs) are perhaps the most studied and well understood DM candidates emerging from BSM~\cite{Arcadi:2017kky,Bertone:2004pz,Jungman:1996}, but unfortunately they have not been detected yet. Specific models, as well as effective ones, have been considered to explore the DM particle properties. Along this line, the restrictions for strongly interacting DM were considered in Ref.~\cite{PhysRevD.41.3594}. 
In addition, DM self-interaction has been considered following the same approach in Refs.~\cite{1992ApJ...398...43C,PhysRevLett.84.3760}. That DM could be charged (millicharged)~\cite{Gould:1989gw,Davidson:2000hf,Dubovsky:2003yn} has also been considered among these phenomenological possibilities, or that it could have an electric or/and magnetic dipole moment~\cite{Heo:2009vt,Heo:2009xt,Masso:2009mu,Profumo:2006im,PhysRevD.70.083501,Barger:2010}, which we shall consider here. 

Observation of large structure formation suggests that DM is made of non-relativistic particles which mainly interact on a gravitational way with SM particles. Non-gravitational interactions might exist but they should be very weak in order to generate the observed large scale structures. Therefore, the DM coupling to photons is assumed to be negligible~\cite{Heo:2009xt}. However, although DM particles are assumed as chargeless, they could be coupled to photons through radiative corrections in the electric ($
D$) and magnetic ($M$) dipole moment~\cite{Masso:2009mu}. Then, we assume DM as fermionic WIMPs endowed with a permanent electric and/or magnetic dipole moment~\cite{PhysRevD.70.083501}.

 Similarly to other WIMP candidates, Dipolar Dark Matter (DDM) particles might be detected either through direct and indirect methods. In the former, WIMPs would be detected by measuring a nuclear recoil produced in their elastic collision with the detector nuclei as target in the laboratory frame~\cite{Heo:2009vt,Heo:2009xt,Masso:2009mu}. Examples of these experiments are CRESST~\cite{Petricca:2017zdp,CRESST:2019jnq}, XENON1T~\cite{Alfonsi:2016mzh,Essig:2017kqs}, CDMS~\cite{Agnese:2015ywx,Witte:2017qsy}, DAMA~\cite{Bernabei:2019uxe,Kang:2019fvz,Baum:2018ekm} and COGENT~\cite{Aalseth:2012if,Davis:2014bla,Aalseth:2014eft}. Besides, indirect methods allow us to detect a WIMP through the observation of secondary products emitted due to annihilation of $\chi\bar{\chi}$ pairs across the galactic halo or inside the Sun and the Earth, where they could have been gravitationally trapped. In this annihilation some kind of radiation would be emitted, such as: high energy photons (gamma rays), neutrinos, electron-positron and proton-antiproton pairs, among other particles. Some examples of the experiments devoted to indirect detection of DM are HAWC (High Altitude Water Cherenkov)~\cite{Abeysekara:2013tza}, FERMI-LAT~\cite{PhysRevD.95.103005,Ackermann:2015tah,Kong:2014haa}, AMS EXPERIMENT~\cite{Xu:2020zfh}, GAMMA-400~\cite{Egorov:2020cmx}, MAGIC~\cite{Doro:2017dqn}, HESS-II~\cite{HESS-II}, CTA~\cite{Acharyya:2020sbj}, and some others.

Furthermore, possible signatures of DDM could arise in some cosmological grounds. Firstly,  like any other WIMP, the cosmic relic abundance due to these DDM particles would have been formed in the early Universe owing to non-equilibrium  thermal decoupling when the pair-annihilation rate dropped below the expansion rate of the Universe.  
To analyze the implications of the annihilation of fermionic DM considering a WIMP with non vanishing magnetic ($M$) and electric ($D$) dipole moments is the goal of this paper. 
Stringent constraints on $\sigma_{ann}$ are obtained from the measurements in high energy experiments and astrophysical sources. 

On one hand, cosmological observations provide weaker constraints to $\sigma_{ann}$ since scattering processes involving DM affect the thermodynamics of cosmic plasma due to injection energy and entropy to this one at the early Universe. Furthermore, another constraint on DDM model can be derived from recent measurements of the temperature anisotropies of the cosmic background radiation and the current relic abundance.

Then, in section~\ref{sec:theoretical} we set up the theoretical framework behind the sort of DM considered here. Interaction $\chi \bar{\chi} \gamma$, and a hierarchy of DDM are given in section~\ref{sec:ddmframework}, moreover in subsection~\ref{subsec:effectiveLagrangian} we  introduce the effective Lagrangian describing the interaction between DDM and photons, as well as the calculation of the thermally averaged cross section corresponding to the annihilation process $\langle\sigma_{ann}v_{rel}\rangle$. 
Afterwards, in section~\ref{sec:results} we present our main results, namely constraints on the magnetic and electric dipole moments, and the DM mass are imposed by requiring the residual abundance and the cross section to be consistent with measurements of the temperature anisotropies of the cosmic background radiation, subsections~\ref{sub:ResultsCMB} and~\ref{sub:Relic}.  Finally, we give our conclusions in section~\ref{sec:conclusions}. 
\section{DDM theoretical framework or DM annihilation in the early Universe}\label{sec:theoretical}
Within the DDM framework $\chi\bar{\chi}$ pairs are able to annihilate into two photons through processes corresponding to the Feynman diagrams shown in Figure~\ref{fig:figure_1}. There are other relevant annihilation processes such as $\chi\bar{\chi}\rightarrow\gamma\,Z^0$ and $\chi\bar{\chi}\rightarrow\gamma\,H^0$. However in this work we assume that the $\gamma\gamma$ channel is the most relevant in the cosmological scenario~\cite{Bringmann:2012}.

According to~\cite{Bringmann:2012}, the spectrum of secondary photons produced by annihilation is homogeneous and has a cutoff at $E_\chi=m_\chi$ where as it barely depends on $m_\chi$ and takes the same form for any channel, therefore we can assume that photons are monochromatic.

At some point, annihilations become fairly unlikely due to the cosmic expansion and the DM species goes out of equilibrium and freezes in Ref.~\cite{Kolb:1990}. This out-of equilibrium process leaves behind a DM cold relic that barely interacts with itself or any other particle except gravitationally~\cite{Bergstrom:2000}. 

For such a thermal particle with a weak-scale mass that annihilates through the $s$-channel, the relative density corresponding to its relic abundance can be inferred from different cosmological observations~\cite{Planck:2019nip,Kolb:1990,Bertone:2005}. In particular, it is well known that the peak-structure of the CMB anisotropies is sensitive to the total amount of DM in the Universe at recombination time~\cite{article}. Therefore, in accordance with the recent precise measurements of these temperature anisotropies made by Planck, the required total amount of cold DM at that epoch must be $\Omega_{CDM} h^2=0.1198\pm0.0012$~\cite{Planck:2019nip}. Besides, this relative density can be computed through the asymptotic Boltzmann equation governing the thermodynamics of massive DM species during annihilation at the early Universe. In this process, as the more efficient is the annihilation process -for larger \sigmav- the smaller would be the left-over DM after decoupling. Thus, this residual quantity is closely related to the thermally-averaged cross section and bounds to the relative density give rise to constraints on the cross section via the following relation~\cite{PhysRevD.79.015014,Kolb:1990}
\begin{equation}
\label{eq:densywimp}
\Omega_{CDM} h^{2} \approx \frac{3\times10^{-26} \textmd{cm}^{3}/\textmd{s}}{\langle\sigma_{ann} v_{rel}\rangle},
\end{equation}
with $v_{rel}$ being the relative velocity.

In this way, the energy density of residual DDM particles is fixed by $\sigma_{ann}v_{rel}$. Note that the previous equation is in agreement with the description above, smaller effective annihilation sections correspond to much higher residual densities. It is worth mentioning that, the value of $\Omega_{CDM}h^2$ shown above is an upper bound for the energy density of the DM relic. In a more realistic scenario, more than one DM species should be considered and therefore their overall energy density must not overpass such value.

In addition, annihilation of sufficiently light DDM particles might have an effect of energy and entropy injection into the cosmic plasma nearby the recombination epoch that results in an effective increase in the free electron fraction leading to a modification in the structure of the CMB spectrum~\cite{Padmanabhan:2005,Kolb:1990}.

Nonetheless, the cosmological bounds are relevant by providing information about the features of DM in a quite different regime. For example, it is well known that if DM couples to gauge bosons, a resonance can be created which is amplified by low-velocity DM at late times in typical astrophysical environments, and it is able to increase the cross section by orders of magnitude, this effect is know as the Sommerfeld enhancement~\cite{Lattanzi:2008}. For that reason, even though constraints from high energy phenomena are stronger than those from cosmological observations, the latter are important to shape the features of DM at the early Universe.

On the other hand, beyond the cosmological scenario, gamma rays provide a valuable piece of astronomical evidence for studying DM annihilation at local scales,  since these photons are not deflected by intermediate magnetic fields between the source and the Earth, therefore the line of sight points towards the target where they are created. This allows us to look for gamma-ray signatures not only in our neighborhood of the galaxy, but also in distant objects such as satellite galaxies, the Milky Way, or even clusters of galaxies. Another advantage of the use of gamma-rays is that, in the local Universe, they do not suffer attenuation and, therefore, they retain the spectral information unchanged on Earth~\cite{Funk:2013gxa}. These advantageous features of gamma rays make the HAWC observatory appealing for studying observational signatures of DM candidates in general and specifically DDM.

As it was mentioned above, this work is only considering annihilation on the $\gamma\gamma$ channel as we are focusing on the implications this contribution may have over gamma-ray observatories, space observatories as Planck and the relic density, and in the future we are expanding the exploration adding some other interesting channels as $f\bar{f}$. Particularly, the $f\bar{f}$ channel is more relevant for small DM masses, as $\gamma\gamma$ contribution is proportional to $m_{DM}^2$.

\section{Dipolar Dark Matter framework \label{sec:ddmframework}}

\subsection{The interaction \texorpdfstring{$\chi\chi\gamma$}{Lg}}
\label{sub:effectiveLagrangian}

Although DM has zero electric charge it may couple to photons through loops in the form of electric and magnetic dipole moments. In this work, a Dirac fermionic DM candidate with an electric and magnetic dipole moments is proposed. The signal proposed as an indirect detection channel is ultraenergetic gamma radiation through the interaction $ \chi \chi \gamma $. By starting from a general form of the coupling $\chi \overline{\chi} \gamma$  in a SM extension, the annihilation cross section $\sigma_{ann} \equiv \sigma(\chi \overline{\chi} \rightarrow \gamma \gamma)$ is analytically computed.
This type of interaction occurs with a BSM particle. As one knows, the proposal of dark matter candidates using gamma radiation in the final state is presented in several models in literature. Here, as in reference~\cite{PhysRevD.70.083501}, we focus on dipole matter in an effective model whose coupling is given by the electric and magnetic dipole properties of the type
\[
\sigma_{\mu\nu}(M+D\gamma^{5}),
\]
where $\sigma_{\mu\nu}=i\left[ \gamma_\mu,\gamma_\nu\right]/2$ is the commutator of two Dirac matrices, $M$ and $D$ are the magnetic and electric dipole moment respectively, in an effective Lagrangian frame. Although fermionic neutral DM particle is proposed, its properties implies an associated millicharge.

Because the candidate is proposed as a stable elementary particle, $M$ and $D$ just can take values such that dipole moment is greater than $D>3.4\times 10^{-9}\, e\, \textrm{cm}$ with $m_{\chi} < < m_e$, and on the other hand if $m_\chi>>m_p$ and $D>1.8 \times 10^{-12}\, e\, \textrm{cm}$, possible bound states with electron and proton respectively are taken into account~\cite{Taoso_2008}. Finally, our DM candidate is a WIMP, that is, a cold, highly stable and neutral particle.

\subsection{A hierarchy of DDM \label{subsec:hierarchy}}
In the general scenario of the study and search of a DM candidate, the mass range is wide (from some keV to TeV). If we take into account the DDM candidate properties, an specific WIMP mass range can be look with favor and can be analyzed within the framework of the experimental constraints given in Sigurdson et al. through Figure 1~\cite{PhysRevD.70.083501}.
With this in mind, we can consider three scenarios regarding the $M$\&$D$ relation:
I) $D>>M$ that implies $f>1$, II) $M\sim D$ that implies $f\simeq 1$, and III) $M>>D$ that implies $f<1$.

The third scenario ($f<1$) is closely related with CP-violation problem. C. Cesarotti et al.~\cite{Cesarotti:2019} analyze the latest ACME results on the limit of electron dipole moment (EDM) $ D_e=1.1 \times 10^{-29}\; e \,\textrm{cm} $ on frame of several BSM theories. In particular, this result can enhance the mass range of DM candidate at  $O(\textrm{TeV})$, imposing strong constraints on Supersymmetry at the LHC (direct detection). Considering the fact that DDM violates CP symmetry, it would imply an additional source that would shed light on the strong CP SUSY problem. On this paper, scenarios I and II are covered in the following section~\ref{sec:results}.

\subsection{The effective Lagrangian for coupling\label{subsec:effectiveLagrangian}}
The effective Lagrangian for the coupling of a Dirac fermion with magnetic and electric dipole moment with the electromagnetic field is~\cite{PhysRevD.70.083501}
\begin{equation}\label{eq:lagrangiano}
{\mathcal{L}}_{\gamma\chi}=-\frac{i}{2}\overline{\chi}\sigma_{\mu\nu}(M+D\gamma^{5})\chi F^{\mu\nu},
\end{equation}
where $\chi$ denotes the DDM field, $F^{\mu\nu}$ is the electromagnetic tensor, and the coupling form is given as $\sigma_{\mu\nu}(M+D\gamma^{5})$. 

\begin{figure}[h]
	\includegraphics[width=0.5\textwidth,height=0.4\textwidth]{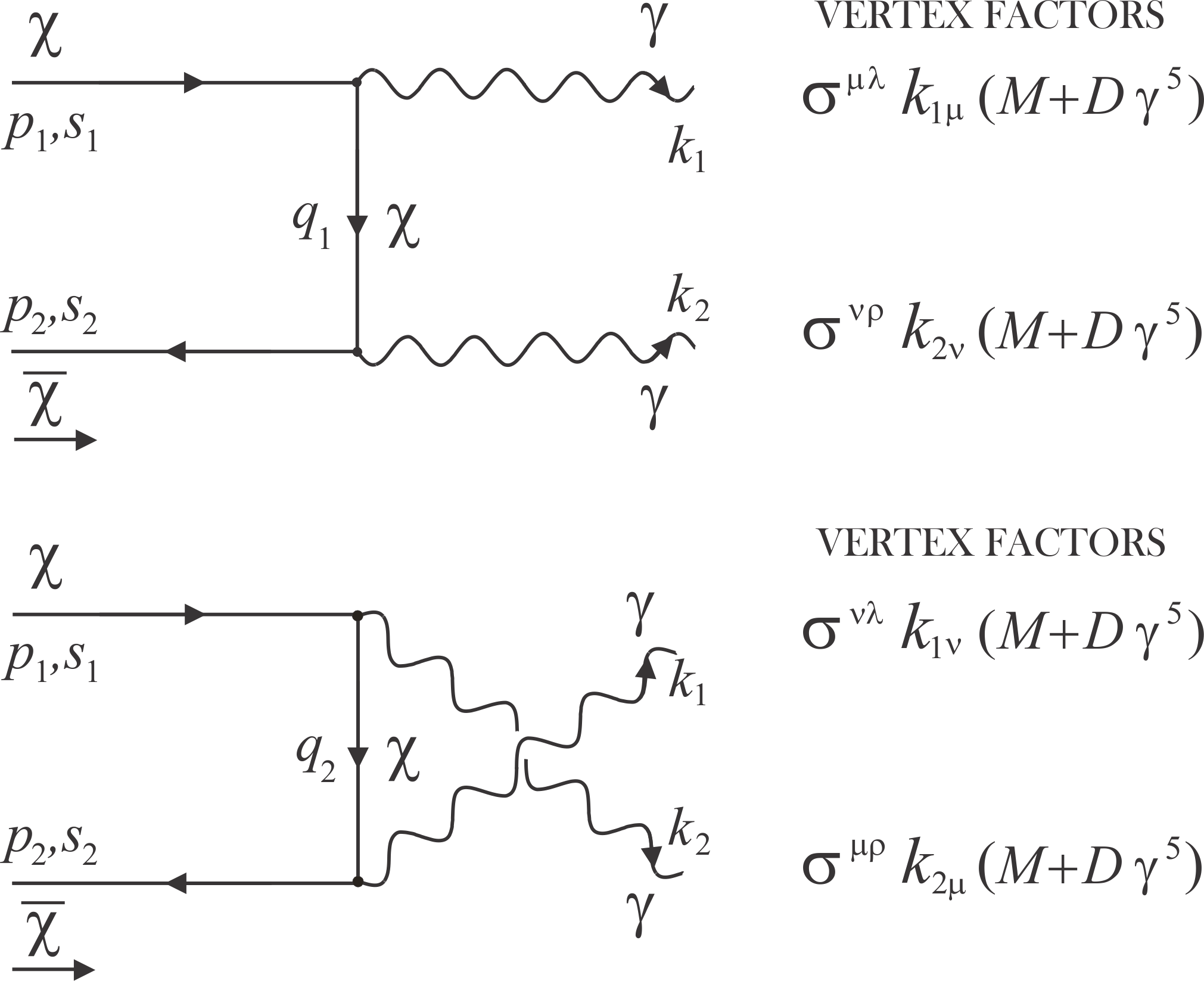}
	\caption{Feynman diagrams for $\chi \overline{\chi} \rightarrow \gamma \gamma$.}
	\label{fig:figure_1}
\end{figure}

For low energies such that $\gamma$-energy and DDM mass relation $E_\gamma/m_\chi$, the photon is blind for $M-D$ difference. In Equation~\eqref{eq:lagrangiano}, $\chi \overline{\chi}$ pairs in the galactic halo or contained in any region of the Universe with high densities (centers of galaxies, clusters of galaxies), can annihilate directly to $\gamma X$, where $X=\gamma,\, Z, H$. In this work we assume that the annihilation of DDM particles is towards two photons through the diagrams shown in Figure~\ref{fig:figure_1}. Annihilations take place mainly through $s$-waves, so $\sigma_{ann}v_{rel}$  is almost independent of the speed and therefore independent of the temperature~\cite{Bergstrom:2000}. %
%

\section{Current constraints on DDM\label{sec:currentconst}}
\subsection{Observational signatures of DDM\label{subsec:cosmological}}
There are multiple experiments currently observing the skies looking for new signals. Some of these expected events can be produced by the annihilation of DM particles in our galaxy, particularly from the galactic center, and a subset of other sources as dwarf galaxies, other known galaxies and galaxy clusters. As it was mentioned previously, HAWC and other gamma-ray observatories are useful in constraining models as we can set upper-bounds on the annihilation cross sections for the $\gamma$-ray production processes and therefore the parameter space, to identify allowed regions that can be consistent with indirect DM detection techniques as well as direct detection, relic abundance limits and collider searches. 

Although HAWC is sensitive to photons from 100 GeV to 100 TeV, it has a maximum sensitivity in the range of 10 to 20 TeV, which makes it sensitive to diverse searches for DM annihilation, including extended sources, diffuse emission of gamma-rays, and the gamma rays emission coming of sub-halos of non-luminous DM. A subset of these sources includes dwarf galaxies, galaxy M31, the Virgo cluster and the galactic center. Likewise, the response of HAWC to gamma rays from these sources has been simulated in several channels of well-motivated DM annihilation ($bb,\,tt, \, \tau \tau,\, W^+W^-$)~\cite{PhysRevD.90.122002}.  By now, this task is out of the scope of this work, nevertheless we plan to resume it in future works.

\subsection{Effective cross section of the annihilation process: \texorpdfstring{$\chi \overline{\chi} \rightarrow\gamma \gamma$}{Lg} \label{subsec:effective}}
We consider the annihilation process $\chi \overline{\chi} \rightarrow\gamma \gamma$ with the same DDM particle as the propagator. Cross section is computed in the frame of center of mass (CM). For this process one have two contributions at low order (see Figure~\ref{fig:figure_1}).

We need to compute $<\sigma_{ann} v_{rel}>$  so that, using the method given by J. D. Wells in Ref.~\cite{Wells:1994qy}, we express $\frac{d\sigma_{ann}}{d\Omega}$ in terms of the Mandelstam variables ($s,t$)
\begin{equation}
\begin{array}{rcl}
\frac{d\sigma_{ann}}{d\Omega} 
&=& \dfrac{1}{128 \pi^{2}} \frac{\sqrt{1-v_{\mathrm{cm}}^{2}}}{m_{\chi} v_{\mathrm{cm}} \sqrt{s}} \times \\ [0.5cm]
&& \times \left[ \dfrac{4\left(D^{4}+M^{4}\right)\left( m_{\chi}^{4}\left(s^{2}+10 s t+6t^{2}\right)-4 t m_{\chi}^{2}(s+t)^{2}+m_{\chi}^{8}+t^{2}(s+t)^{2}\right)}{\left(m_{\chi}^{2}-t\right)\left(-m_{\chi}^{2}+s+t\right)}\right. \\ [0.5cm]
&&\left. + \dfrac{8 D^{2} M^{2}\left(-3 m_{\chi}^{4}\left(s^{2}+2 s t-2 t^{2}\right)+4 t m_{\chi}^{2}\left(s^{2}-t^{2}\right)+4 m_{\chi}^{6}(s-t)+m_{\chi}^{8}+t^{2}(s+t)^{2}\right)}
{\left(t-m_{\chi}^{2}\right)\left(-m_{\chi}^{2}+s+t\right)}  \right],
\end{array}
\label{eq:mandelstam}
\end{equation}
where
 $|\textbf{p}_{1}| = \frac{m_{\chi} v_{\mathrm{cm}}}{\sqrt{1-v_{\mathrm{cm}}^{2}}}$, 
$|\textbf{p}_{3}| = \frac{\sqrt{s}}{2}$,
$(E_{1}+E_{2})^2 = s$,
$v_{cm} = \dfrac{v_{rel}}{2}$.
Finally, in order to get the average of the thermal distribution of the WIMPs, we need to compute \sigmav over all the phase-space variables. In that way, we get \sigmav, needed to carry out further thermal analysis such as computing the relic abundance. 

Since DM mean velocity is almost vanishing -hence the velocity dispersion-, we must work within the non-relativistic limit. 
Then, starting of Equation~\eqref{eq:mandelstam}, where it is possible to use the method described in Ref.~\cite{Wells:1994qy},  \sigmav is given as
\begin{equation}
\begin{array}{rcl}
\left\langle\sigma_{ann}v_{rel} \right\rangle & = & \tilde{c}_0  \, \, m_{GeV}^2 \bigg[6(M_{16}^4+6M_{16}^2D_{16}^2+D_{16}^4)\\
&&+(3M_{16}^4+2M_{16}^2D_{16}^2+3D_{16}^4)\left\langle v_{rel}^2\right\rangle \bigg] \textrm{cm}^3 \textrm{s}^{-1},
\end{array}
\label{eq:sigmaprom}
\end{equation}
 where $\tilde{c}_0=1.71423\times 10^{-30}$, $m_{GeV}=\frac{m_\chi}{\textrm{GeV}}$, and both the magnetic and electric dipole moments have been normalized to be dimensionless: $D, M\rightarrow D_{16}=D/10^{-16}, M_{16}=M/10^{-16}$. On the other hand, using the relation $\langle \sigma_{ann}v_{rel}\rangle \approx  a_{rel} + b_{rel} \langle v_{rel}^2  \rangle=a_{rel}+\dfrac{6b_{rel}}{x}$ given in~\cite{Cannoni:2015wba}, $\langle v_{rel}^2  \rangle=\dfrac{6}{x}$. Now, we can rewrite the predicted \sigmav in \eqref{eq:sigmaprom} as,
\begin{equation}
 \langle\sigma_{ann}\,v_{rel}\rangle 
 = \tilde{c}_0 \, m_{GeV}^2 \, M_{16}^4 H(f,x),
 \label{eq:termalsigma}
\end{equation}
where  $x=\frac{m_\chi}{T}$ is a dimensionless quantity ($T$ is  decoupling temperature), which in the non-relativistic limit $x >> 1$ (or $T<< m_\chi$), $f\equiv\frac{D_{16}}{M_{16}}$ is the dimensionless parameter that corresponds to the ratio of electric to magnetic dipole moments respectively), and the dimensionless function $H(f,x)$ has the following form:
\begin{equation}
 H(f,x) = 6(1+6f^2+f^4) + \frac{6}{x}(3+2f^2+3f^4).
\end{equation}
 We set $x$ to the magical number $x=\frac{m_\chi}{T}\sim 22$, which is a typical value for WIMPs~\cite{PhysRevD.76.103524}.
In this way, the theoretical parameter-set we shall use onward is $\{m_\chi,M_{16},f\}$. 

Notice that \sigmav  increases either if $m_\chi$ and $M_{16}$ do it, which implies that  if $m_\chi$ or $M_{16}$ increase, $\chi\overline{\chi}$ pairs annihilate more efficiently. Since the  factor
$H(f,x)$  is around $50$ when $f\sim 1$,  $m_\chi$ and  $M_{16}$  control the order of magnitude of \sigmav.

\section{Constraints on the DDM parameters space according to Planck\label{sec:results}}
In this section we determine a parameter-subspace of DDM models that is consistent with some cosmological constraints for the thermally averaged DM annihilation cross section  derived from Planck measurements of the temperature anisotropies of the cosmic background radiation. Firstly, in subsection~\ref{sub:ResultsCMB}, we
consider phenomenological constraints in the $f_e$\sigmam plane (where $f_e$ a non perfect absorption efficiency is assumed) derived by Masi et al.~\cite{Masi:2015} and by Kawasaki et al.~\cite{Kawasaki:2015}, in order to infer the corresponding implications within our specific model and derive the allowed region of parameter space accordingly. The resulting bounds on the parameters will be taken as a prior assumption in our further statistical analysis carried out in the next section. 
Secondly, in~\ref{sub:Relic} we determine the projected posterior probability distribution (PPD) for $m_\chi$, such that our prediction of the relative density of the cold DM relic $\Omega_{CDM}h^2$ is consistent with the most recent measurements according to Planck~\cite{Planck:2019nip}. For that purpose, we sample the DDM parameter space using a three dimensional grid-mesh in order to compute the goodness-of-fit estimator $\chi^2$ associated to the previously mentioned data-set. 

 \subsection{Bounds from the relative density of DM relic\label{sub:Relic}}
 In this subsection we derive bounds on the dipole moments and the mass of DDM from requiring that the predicted cold relic to be in accordance to the latest measurement by Planck. In the first part, we consider the whole three dimensional parameter space described at the beginning of this section. It is usual to fix the electric to magnetic dipole moment ratio to $f=1$ under the argument that $M_{16}$ and $D_{16}$ have the same order. However, even if  $f\sim 1$, here we show that theoretical curves in the $\Omega_{CDM}h^2-m_\chi$ plane are importantly sensitive to variations of $f$. 
 
 \subsubsection{Scenario I: $D>>M$} 
\begin{figure}[ht]
 	\centering
	\includegraphics[width=0.5\textwidth,height=0.4\textwidth]{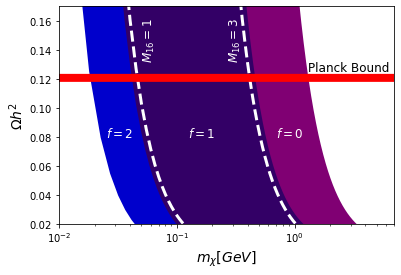}
	\caption{ Blue, purple and magenta regions correspond to the theoretically predicted regions of the DM relative density for  $f=2,1,0$ respectively. Within each region  $M_{16}$ runs between $[1,3]$. Graph shows the residual abundance~\eqref{eq:densywimp} for $D_{16}=3$.}
\label{fig:figure_6}
\vspace{15mm}
\end{figure}

 In this section we study the regions of the DDM space of parameters for a wide ranges of values
of $m_\chi $. Naturally, the purpose is to identify the regions of highest likelihood in accordance to the latest bounds to the DM relative density inferred from Planck. Before that, with the aim of getting an idea of the degree of sensitivity of $\Omega_{CDM}h^2$ to variations of the theoretical parameters, we explore the predicted $\Omega_{CDM}h^2-m_\chi$ curves for different models. 
 Figure~\ref{fig:figure_6} illustrates the effect of varying the dipole moments parameters $M_{16}$ and $f$ over $\Omega_{CDM}h^2$ as function of the DM particle mass. More specifically, three classes of $\Omega_{CDM}h^2(m_\chi)$ curves are shown which have a common value of $f$ (associated with a given color). Each class contains curves of models corresponding to different values of $M_{16}$ within a fixed range of values of order 1. For a given $M_{16}$, the effect of varying $f$ is clear, as it increases the $\Omega_{CDM}h^2-m_\chi$ curves shift to smaller $m_\chi$.
 
 As a consequence, the values of $m_\chi$ picked by the data for DDM with non vanishing electric dipole moment (for a given value of $M_{16}$) are smaller than those for DM holding only magnetic dipole moment. In other words, light DDM particles holding electric dipole moment are able to annihilate at the same rate than heavier particles with $f=0$. In addition, there exists an overlap of curves of different classes, which is an indicator of a possible degeneracy between the parameters.  

 Now let us proceed to describe the procedure used to sample the DDM parameter space according to the data. We calculated the $\chi^2$ statistical estimator corresponding to the most recent measurement of the energy density parameter for the relic abundance of DM from Planck data-sets shown in the Table~I.
\begin{table}
\begin{tabular}{|c|c|c|c|}
\hline
&Plik&CamSpec&Combined\\
\hline
$\Omega_{CDM}h^2\pm1\sigma$& $0.12\pm0.0012$& $0.1197\pm0.0012$&$0.1198\pm0.0012$\\ 
\hline 
\end{tabular}
\caption{Cold DM relative density from Planck data-sets.}
\label{table:1}
\end{table}
For this purpose, the corresponding theoretical prediction is related to the thermally averaged cross section computed previously in~\ref{eq:densywimp}.

We can assume, as a fair approximation, a normal likelihood distribution for $\Omega_{CDM}$ considering a flat prior. Thus, based on the Bayes theorem, the posterior probability of a model associated to a point in parameter space $(m_\chi,M_{16},f)$ to describe a measurement of $\Omega_{CDM} h^2$ reads
\begin{eqnarray}\nonumber
P( \Omega_{CDM}^{(th)}| \Omega _{CDM}^{(ob)}) \sim e^{-\chi^2}\qquad\text{where}\qquad 
\chi^2(m_\chi,M,f)&=&\frac{  \left(\Omega _{CDM}^{(th)}(m_\chi,M,f)- \Omega _{CDM}^{(ob)}\right)^2  }{\sigma^2}  
\end{eqnarray}
where $\sigma$ is the observational error and $\Omega_{CDM}^{(ob)}$ is the best-fit central value of the Planck collaboration estimation. 
By computing numerically the $\chi^2$ as described below, we sampled the parameter's space of the DDM model and used the bound found in section~\ref{sub:ResultsCMB} as a prior. 

\begin{figure}[ht]
\includegraphics[width=0.65\textwidth]{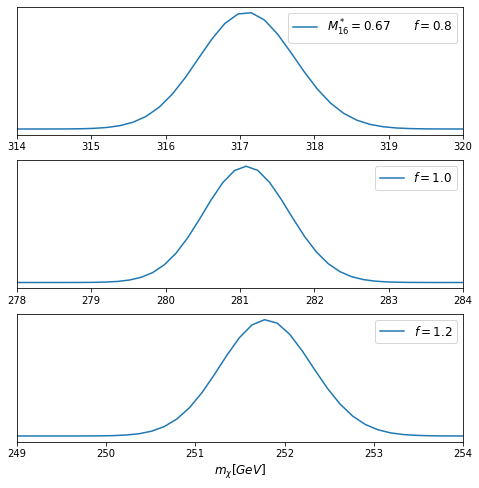}
\caption{1D projection posterior distributions for the DM mass for the following values of the electric/magnetic dipole moments ratio $f=0.8,1.0,1.2$ and magnetic dipole moment equal to the cutoff $M^*_{16}=0.67$ established by CMB-Planck data.}
\label{fig:Posteriors_m}
\end{figure}

The main result of this part is presented in Figure~\ref{fig:Posteriors_m}. In there, various marginalized posterior distributions for some values of the $m_\chi$ parameter are shown for fixed $M_{16}<M_{16}^*$ below the maximum value of the dipole moment consistent with CMB constraints found in~\ref{sub:ResultsCMB} and different values of $f$. Consistently with the analysis made at the beginning of this subsection, the estimation of $m_\chi$ shifts to lower values as $f$ is raised.   
%

\subsubsection{Scenario II: Equal electric and magnetic dipole moments \texorpdfstring{$f=1$}{Lg}}
Now let us consider that $D_{16}=M_{16}$ under the assumption that both electrical and magnetic dipole moments are the same order of magnitude. Therefore, the annihilation effective cross section expression  by the relative thermally averaged speed for the process $\chi \overline{\chi} \rightarrow \gamma \gamma$  is: 
\begin{equation}\label{eq:sigmaprom3}
\begin{split}
\left\langle\sigma_{ann}v_{rel} \right\rangle = &
48 \tilde{c}_0 m_{GeV}^2    D_{16}^4 \bigg[1+ \frac{1}{x} \bigg]\textmd{cm}^3 \textmd{s}^{-1} .
\end{split}
\end{equation}
We are taking into account the upper limit for dipole moments, $D_{16}=M_{16} \leq 3$, reported by K. Sigurdson et al.~\cite{PhysRevD.70.083501} and the dimensionless quantity for the WIMPs, $x \cong 22$. 
 Note that  $\left\langle\sigma_{ann}v_{rel} \right\rangle$ has the order of magnitude corresponding to the total annihilation cross section for a generic WIMP, which is usually set as 
$\left\langle\sigma_{ann} v_{rel}  \right\rangle_{D_{16}=3} \simeq 3 \times 10^{-26} \, \textrm{cm}^3 \textrm{s}^{-1}$~\cite{Steigman:2012nb}.

On the other hand,  we can relate Equation~\eqref{eq:termalsigma} with~\eqref{eq:densywimp} in order to predict the value of the residual density of cold DM $\Omega_{CDM} h^2=0.12$, for a whole set of 
parameters for the DDM model. As a consequence, if we assume that DDM particles constitute the whole amount of cold DM in the Universe and that $f=1$, then the mass of the DDM particle depends on the values of the dipole moment. In specific,  for electrical dipole moments below  $0.25\times10^{-16}e \textrm{cm}$ and masses above $10\,\textrm{GeV}$ it can always be possible to predict the relic abundance inferred by WMAP (see Figure 2). In this particular case in which  $f=1$ is assumed, the upper bound for the dipole moment according to Planck-CMB measurements is strongly raised to $M_{16}=M_{16}^*= 0.67$.
\begin{figure}[ht]
\centering
	\includegraphics[width=0.65\textwidth,height=0.45\textwidth]{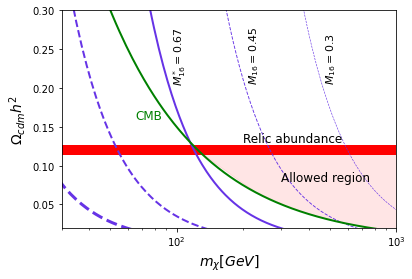}
	\newline
	\caption{The DM relative energy density. The red stripe corresponds to the allowed region according to the bound for the relative density of the relic of DM inferred from CMB measurements by Planck. Each purple line corresponds to the theoretical prediction of the relative density as a function of $m_\chi$. Different dashed lines correspond to several values of the magnetic dipole moment assuming that $M_{16}=D_{16}$.}
		\label{fig:figure_5}
	\end{figure}
 Even though candidates with  $M_{16}>M_{16}^*$ can be compatible with the relic abundance observations for a range of masses, they are excluded by the Planck-CMB measurements.  Particle masses with dipole moment $M_{16}\sim M_{16}*$ (thick purple line in  Figure~\ref{fig:figure_5}) are very restricted and lay in a narrow range around $m_\chi^*\sim 500$ GeV in order to be consistent with relic abundance observations. Although particles of this kind with masses $m_\chi<m_\chi^*$ are allowed by CMB measurements, their thermally averaged annihilation cross section would be insufficient in order to reduce the primeval DM abundance to give the DM  relative density observed today. On the other hand, more massive candidates with the same value of $M_{16}$ are excluded by the CMB observations. Notice that the allowed range of masses is very sensitive to variations of $M_{16} \gtrsim M_{16}^*$. A slight decrease in $M_{16}$ (from $M_{16}^*$ to $M_{16}=0.17$) shifts the $\Omega_{CDM}h^2$ curve almost an order of magnitude towards larger masses. Additionally, models within that range are able to consistently predict the relic abundance but only making up a fraction larger than the green line. 
 
 In conclusion, on this final combined analysis, by considering both the relic abundance and the CMB constraints, a host of models with low masses and large $M_{16}$ are excluded. 

\subsection{Constraints on the DDM parameters from measurements of the Temperature Anisotropies of the CMB by Planck\label{sub:ResultsCMB}}
In section~\ref{sec:theoretical} 
we have already explained how some observable signatures of DDM might appear in the features of the CMB. On one hand, let us recall that the overall DM abundance at recombination strongly determines the shape of the anisotropies of the CMB. On the other hand, DM annihilation injects energy to the  electron-photon gas during such epoch, and consequently, the location and shape of the peaks of the CMB spectrum provide information about the magnitude of the annihilation cross section and the mass of the DM particle~\cite{Kolb:1990}.

Firstly, let us describe the phenomenological constraints considered here. In~\cite{Masi:2015}, bounds in the $f_e$\sigmam \, plane are derived and these are based on preliminary Planck results which are compatible with a $s$-wave annihilation cross section around $10^{-23}\,\textrm{cm}^3\,\textrm{s}^{-1}$ for $\sim$ TeV DM, in their analysis it is assumed an imperfect absorption efficiency $f_{e}\sim 0.2$~\cite{Madhavacheril:2013}. In despite of being preliminary, such constraints are consistent with  posterior bounds reported in~\cite{Kawasaki:2015} inferred from Planck 2015. In that work, the effects of energy injection to the background plasma due to annihilations occurring at higher red-shift are simulated by using the methods established in~\cite{Kanzaki:2009}.
As pointed out by the authors, CMB inferences have some
advantages over cosmic ray ones, namely, CMB constraints do not depend on the DM distribution inside galaxies, which represents a source of systematic errors in cosmic ray experiments.

\begin{figure}[ht]
\centering
\includegraphics[width=0.8\textwidth]{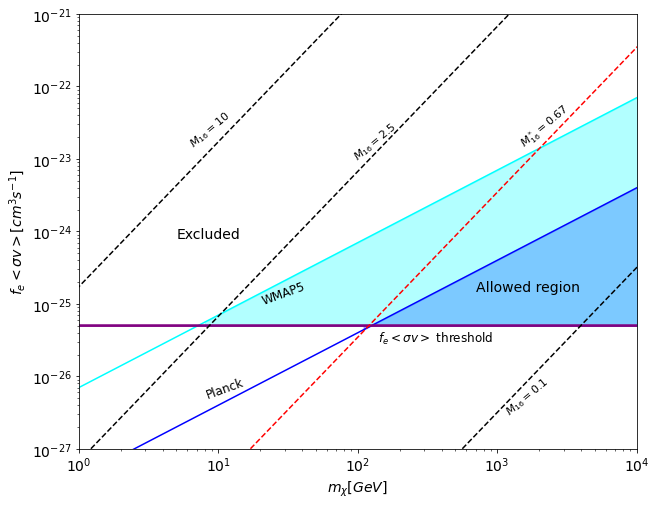}
\caption{Solid lines delimit regions of exclusion in the $f_e$\sigmam\, plane of WIMPs annihilating to photons according to CMB measurements. Dashed lines correspond to theoretical predictions within the DDM model for $f=1$ and $M_{16}=0.117,0.250,0.670,10.0$.}
\label{fig:PlanckBound1}
\end{figure}

Now, let us analyze how constraints from CMB-Planck reduce the domain of the specific functional form of \sigmav\, computed in~\ref{subsec:effective}.
Within the $f_e$\sigmam\, plane, the region laying
below the dark-blue solid line in Figure~\ref{fig:PlanckBound1} corresponds to the allowed models by the mentioned data-set. In there, the theoretical model proposed, gray dashed lines, show  the $f_e$\sigmav\, as function of $m_\chi$ for  $M_{16}=0.117,0.250,0.670,10.0$ and $f=1$. For a model with a given $M_{16}$ there exists an upper bound $m_{up}$, that is, particles with masses above $m_{up}$ are excluded by CMB-Planck. The solid light-blue line in the same plane represents the WMAP5 constraint~\cite{Barreiro:2008pn}. Notice that the upper bound for the mass from this data-set is weaker by more than one order of magnitude in comparison to the one corresponding to Planck. 

Figure~\ref{fig:PlanckBound1} also shows the lower limit for \sigmav, in order not to overpass the observed relic of DM in the Universe (purple solid line). This bound on the cross section gives rise to a lower bound $m_{low}$ on $m_\chi$ for models within the allowed region (dark blue in Figure~\ref{fig:PlanckBound1}).

  For a given dashed line (associated to a single value of the dipole moment), only those values of $m_\chi$  for which the bit of line falling inside the blue region are allowed simultaneously by Planck and the relic abundance measurement. As it can be noticed for the $f=1$ case in Figure ~\ref{fig:PlanckBound1}, there is a cutoff ($M_{16}^*\sim 0.67$ for $f=1$) above which DDM models are excluded. For dipole moments within this allowed threshold, the relic abundance bound provides a lower bound $m_{low}$ satisfying the following relation (in general for any value of $f$),
  \begin{eqnarray}
\langle\sigma_{ann}v_{rel}\rangle_{relic}&=&\langle\sigma_{ann}v_{rel}\rangle^{th}(m_{low},M_{16},f)\\
2.5\times10^{-25} &=&  \tilde{c}_0  \left(\frac{m_{low}}{GeV}\right)^2 M_{16}^4 H(f,22),
  \end{eqnarray}
which leads to 
\begin{equation}\frac{m_{low}}{GeV} = \left(\frac{19.54}{M_{16}}\right)^{2}\left(\frac{1}{H(f,22)}\right)^{1/2} . 
\end{equation} 

Constraints from CMB measurements can be fitted by the following functional form:
  \begin{equation}
f_e\langle\sigma_{ann}v_{rel}\rangle^{Planck}=(4\times 10^{-28}\textrm{cm}^{3}\textrm{s}^{-1}) m_{GeV}.
  \end{equation}
In a similar way as the relic abundance bound provides a lower bound $m_{low}$, the CMB-Planck constraint provides an upper bound for the DM particle mass $m_\chi$ within models with $M_{16}<M_{16}^*$, which satisfies 
${\langle \sigma_{ann}v_{rel} \rangle}^{th}(m_{up},M_{16},f)-(4\times 10^{-28} \textrm{cm}^{3} \textrm{s}^{-1})\,\, m_{GeV}=0$.

 This one implies that
\begin{equation}\label{eq:upperm}
\frac{m_{up}}{GeV}=\left(\frac{5.84}{M_{16}}\right)^{4}\frac{1}{H(f,22)}.
\end{equation}
\begin{figure}[ht]
\centering
\includegraphics[width=0.6\textwidth]{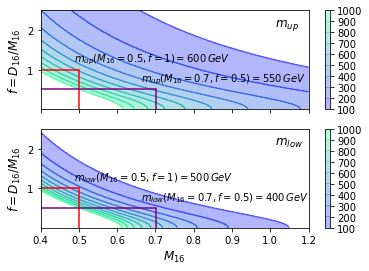}
\caption{Level-contours for $m_{up}(M_{16},f)$(top) and $m_{low}(M_{16},f)$(bottom). Each solid line corresponds to the upper and lower limits of DDM mass respectively for a variety of models with different values of dipole electric and magnetic moments  according to Planck.}
\label{fig:PlanckBound2}
\end{figure}

In addition, for the more general case in which $f$ is a free parameter, the top  and bottom panels in Figure~\ref{fig:PlanckBound2} illustrate the level-contours for functions $m_{up}(M_{16},f)$, and $m_{low}(M_{16},f)$ respectively, corresponding to the upper an lower bounds for models with different magnitudes of electric and magnetic dipole moments. In  Figure~\ref{fig:PlanckBound2}, it is illustrated how both bounds for two chosen models can be determined. Both of them are representative, the first one (orange label) with $M_{16}=0.5$ and $f=1$ has equal electric and magnetic dipole moment close to the cutoff value $M_{16}^*$, while for the second model (purple label) a larger $M_{16}$ is permitted as long the $M_{16}-D_{16}$ ratio is reduced. For both models the allowed mass ranges lay around $400-600\,\textrm{GeV}$.   

It is clear from these figures, that the $f$ parameter notably relaxes the constraint on $M_{16}-m_\chi$ plane as expected. When $f=1$ once $M_{16}$ is fixed, the cutoff for $m_\chi$ is uniquely determined. In contrast, when $M_{16}$ is fixed while $f$ is free to vary, a whole range of masses are allowed. 

\section{Conclusions\label{sec:conclusions}}
In this work we studied the DM annihilation, considering it as a neutral particle with non vanishing magnetic and/or electric moments.  Therefore, we studied this candidate with  $M\sim D$ laying below the upper bound of $3 \times 10^{-16}\, \textrm{e cm}$ obtained in previous works~\cite{PhysRevD.70.083501}. The effective annihilation cross section  $\chi \overline{\chi} \rightarrow \gamma \gamma$  was analytically computed  from first principles.
 In addition, we restricted the parameters space involved in the thermally averaged annihilation cross section in order to be consistent with cosmological data, such as the relative density of the DM relic abundance and measurements of the temperature anisotropies of the cosmic background radiation. Then, we considered the model-independent constraint in the $f_e$\sigmam\, plane derived in~\cite{Kawasaki:2015}, and analyzed its implications over the DDM model parameter space. Firstly, by imposing these CMB bounds and fixing  $M_{16}$, $f$ values, there exists an upper bound for DDM particle mass, $m_{up}$. Moreover, in order to $m_\chi$ to be consistent with measurements of the DM relative density today, a lower bound $m_{low}$ is imposed. The resulting allowed mass range lies around $10^2\,\textrm{GeV}$. We also demonstrated that if $f$ is taken as a free parameter (while remaining order one), then the allowed range of $M_{16}$, consistent with the CMB data-set, becomes wider. As a second step, we analyzed the CMB constraint implications and the relic abundance measurement as well.  By combining both priors, we found the upper cutoff $M_{16}^*=0.67$ for the magnetic dipole moment (when $f \sim 1$). Afterwards, we estimated the projected posterior distributions for $m_\chi$ taken several $f$ values, and by fixing the magnetic dipole moment $M_{16}$ to a value below $M_{16}^*$ in accordance to the CMB prior.

\begin{acknowledgments}
 This work has been partially supported by CONACYT-SNI (M\'exico).
\end{acknowledgments}

%

\end{document}